# Characterization of atmospheric pressure $H_2O/O_2$ gliding arc plasma for the production of OH and O radicals

N. C. Roy, M. G. Hafez[a], and M. R. Talukder

*Department of Applied Physics and Electronic Engineering, University of Rajshahi, Rajshahi 6205 Bangladesh.*

[a] *Department of Applied Mathematics, University of Rajshahi, Rajshahi 6205 Bangladesh.*

Corresponding author: neporoy@gmail.com

**Abstract**

Atmospheric pressure $H_2 0/O_2$ is generated by a 88 $Hz$, $6kV$ AC power supply. The properties of the produced plasma are investigated by optical emission spectroscopy (OES). The relative intensity, rotational, vibrational, excitation temperatures and electron density are studied as function of applied voltage, electrode spacing and oxygen flow rate. The rotational and vibrational temperatures are determined simulating the $OH(X^2\Pi(v' = 0) \rightarrow A^2\sum^+ v'' = 0))$ bands with the aid of *LIFBASE* simulation software. The excitation temperature is obtained from the $CuI$ transition taking non-thermal equilibrium condition into account employing intensity ratio method. The electron density is approximated from the $H_\alpha$ Stark broadening using the Voigt profile fitting method. It is observed that the rotational and vibrational temperatures are decreased with increasing electrode spacing and $O_2$ flow rate, but increased with the applied voltage. The excitation temperature is found to increase with increasing applied voltage and $O_2$ flow rate, but decrease with electrode spacing. The electron density is increased with increasing applied voltage while it seems to downward trend with increasing electrode spacing and $O_2$ flow rate.

## 1. Introduction

In the recent years, water plasmas have drawn much attraction due to its distinctive features of high oxidation-reduction capacity, high enthalpy and above all environment friendly. Due to these features, the water plasmas are applied [1] in environment, biology, hazardous waste treatment, coal gasification and so on [2, 3]. Kim *et al.* [4] carried out experiment to study the performance of polychlorinated biphenyles waste treatment using $H_2 0$ steam plasma and found that the steam plasma process yield better efficiency for converting hazardous waste to energy than the conventional burning and air plasma processes. Ni *et al*[5] characterized steam dc plasma torch at atmospheric pressure employing electrical and optical emission spectroscopic (OES) diagnostic techniques. They observed that the restrike mode is responsible for fluctuations and instabilities of the steam plasma. Zhu *et al.* [6] demonstrated the gliding arc non-equilibrium plasmas by flowing air at atmospheric pressure. They studied the effects of air flow rates on the discharge dynamics and ground state $OH$ distributions. They also noted that the air flow rate strongly affects the discharge ignition, maintenance, and emission intensity. Bruggeman *et al*[7] investigated the electrical and optical properties of dc water discharge plasmas at atmospheric pressure air and noted that such types of plasmas able to produce a large number of $OH, H_2O_2, NO, N_2^+$ and other species. The highly reactive $OH$ and $O$ radicals are recently used[1,8] for the destruction of homicide bacteria, virus and fungus. Therefore, for better understanding of the production and control of the active species in the discharge and their interactions with the living tissues are very important. The production of these active species depends on the conditions of the discharge to generate high concentration excited states of the plasma species that have energies above their ground states and can affect the destruction mechanisms of the living tissues. The addition of $O_2$ in $H_20$ steam discharge can enhance the productions of reactive $OH$ and $O$ radicals through different collision processes. Therefore, it is important to determine plasma parameters (gas temperature, electron temperature, electron density) at different percentage of oxygen in $H_20$ steam plasma under different discharge conditions will help to understand about particle collision processes, plasma reactions and density of the react species in the plasma. OES is a non-perturbing and non-invasive easy-to-use plasma diagnostic technique and widely applied to investigate plasma species and to determine gas temperatures. OES is able to identify plasma species such as free radicals, atomic and molecular species and therefore it reveals the insight of the plasma chemical processes[9]. OES also provides valuable information regarding excited atomic or molecular species that enables us to estimate rotational and vibrational temperatures of the species in non-equilibrium $H_2O/O_2$ gliding arc plasma.

The water $H_20/O_2$gliding arc plasma is characterized by OES method to investigate the production of reactive species in terms of oxygen fraction in the mixture and operating parameters. The main objectives of this work is to gain insight of the reactive $OH$ and $O$ species that influence the plasma reactivity and to obtain better understanding of the production mechanisms of the species concerned. Experimental setup is discussed in section 2, optical characterization is given in section 3, results and discussions are furnished in section 4, and finally conclusion is drawn in section 5.

## 2. Experimental Setup

Figure 1 shows the schematic illustration of the experimental setup. A $65\ mm$ long, $10mm$ inner diameter and $1mm$ thick glass tube is used as reactor chamber. The diameter of the nozzle at the exit point is $2mm$. Two copper electrodes of 10mm long and $1mm$ diameter two copper electrodes are inserted from the opposite sides into the discharge tube. Copper electrodes are intentionally used because the distinct $CuI$ line can be used for the determination of excitation temperature of the plasma. Experimental measurements are carried out from $2-5\ mm$ electrode spacing. A small amount of $H_20$ steam with a constant temperature of $\sim 400\ K$ from the steam generator is fed with a constant flow rate to the plasma reactor as shown in Fig. 1. $O_2$ (purity 99.65%) is fed to the discharge reactor from the gas cylinder. It is noted that the $O_2$ flow rates are varied but the $H_20$ steam flow rate remains constant at $0.5\ lpm$ in the entire experimental investigations. Flow rate of $O_2$ inside the discharge tube is controlled by a gas flow controller ($KIT\ 115P$). The temperatures of $H_20$ and $O_2$ are ~400K and 300K, respectively, at the entrance of the discharge tube while the $H_20/O_2$ mixture is 328K at the tube exit point without discharge. The observed temperature difference is $\sim 5K$ for the $O_2$ flow rate from $2-10\ lpm$. A high voltage (maximum $6kV, \sim 25W$) power supply with a frequency $88Hz$ is applied across the electrodes. The waveforms of the discharge voltage and current are recorded with a voltage probe ($HVP-08$), and a current probe ($CP-07C$) in combination with a digital oscilloscope ($GDS1022$). In order to record the emission spectra of the produced plasma a $200\ cm$ long optical fiber cable is fed to the spectrometer ($USB2000+XR1$). The spectrometer has entrance slit of $25\mu m$, detector wavelength range of $200-850\ nm$ and optical resolution of $1.7nm$. A computer is associated to acquire spectral data.

## 3. Optical Characterization

*Species identification*

Optical emission spectroscopy is a common technique used for species identifications and plasma diagnostics at atmospheric pressure plasmas. Figure 2 shows a typical emission spectrum measured at $5kV$ with $O_2$ flow rate of $4\ lpm$ $(11.11\%\ H_2O)$ for the electrode spacing of $3mm$. The dominant peaks are found for the band of $OH(X^2\Pi(v^{'}=0) \rightarrow A^2\sum^{+} v^{''}=0))$ at $305-320nm$, and $O(4s(\ ^3D)4d \rightarrow 4p(\ ^3P^{\circ}))$ at $777\ nm$, respectively, in the presence of $H_20$. It is well known that $OH$ and $O$ are the most important species for plasma chemistry due to their highly reactive property [10]. The emissions of $CuI$ $(4s(\ ^3D)4d \rightarrow 4p(\ ^3P^{\circ}))$ at $324.32\ nm$ and $(4s(\ ^1D)4d \rightarrow 4s4p(\ ^3P^{\circ}))$ at $326.60\ nm$ lines are observed due to erosion of the copper electrodes in the highly reactive $H_20/O_2$ plasmas. The transition of $H_\alpha(3s \rightarrow 2p)$ and $H_\beta(4d \rightarrow 2p)$ are found at $486.1\ nm$ and $656.3\ nm$, respectively. The emission intensity of $H_\alpha$ line is observed much stronger than that of $H_\beta$ line due to the lower ionization potential of $H_\alpha$ compared to $H_\beta$ and hence more ionization event occurs with dissipation of the same amount of energy. Therefore, the density of $H_\alpha$ can be expected higher. The $H_\alpha$ line will be used for the determination of electron density.

The basic principle of OES method is that the emission intensity from an excited state is proportional to the concentration of the species in that excited state at a particular wavelength[11,12]. At atmospheric pressure, in most cases, the lifetime of the species is large enough where the collisions thermalize the rotational state distribution function with gas temperature $T_g$ and the rotational temperature $T_{rot} \approx T_g$ The optical emission intensity of the species can be written as [9, 13]

$$I_{n^{''}v^{''}J^{''}}^{n^{'}v^{'}J^{'}} \propto N_{n^{'}v^{'}J^{'}} A_{n^{''}v^{''}J^{''}}^{n^{'}v^{'}J^{'}} h\nu_{n^{''}v^{''}J^{''}}^{n^{'}v^{'}J^{'}}, \qquad (1)$$

where $I$, $N$, $A$, $h$, ν, $n$, $v$ and $J$ are the emission line intensity, density of the molecules, Einstein's transition probability, Plank's constant, frequency of transition, quantum numbers of electronic, vibrational and rotational transitions, respectively. The single and double primes indicate the upper and lower levels of transitions, respectively. The $O$ and $OH$ radicals are produced in the gap of the electrodes through electron impact dissociation of $O_2$ and $H_2O$ molecules, as described in reactions $(R1)$ and $(R2)$, respectively. Inserting $T_e = 0.58eV$ as the typical value found in our experiment, in the rate coefficient equation of reactions $(R1)$ and $(R2)$, the production rates of $O$ and $OH$ radicals are $k_1 = 2.59 \times 10^{-13} cm^3 s^{-1}$ and $k_2 = 9.82 \times 10^{-11} cm^3 s^{-1}$, respectively, which indicate that the density of the $OH$ radical is higher than those of $O$, as can be seen from Fig. 3. On the other hand, the kinetic energy of the electron is boosted up by the enhanced electric field due to increased applied voltage, hence the collision frequency increases and the production of $O$ and $OH$ radicals are increased which can be inferred from reactions $(R1)$ and $(R2)$, respectively. The density of $O$ and $OH$ radicals are also increased with increasing $O_2$ flow rate [14].

$$e + O_2 \rightarrow 2O + e \qquad k_1 = 4.2 \times 10^{-11} \exp\left(-\frac{5.6}{T_e}\right)\ cm^3 s^{-1}. \qquad (R1)$$ [14]

$$e + H_2O \rightarrow OH(A) + H + e \qquad k_2 = 10^{-10}(1.68T_e - 1.23T_e^2 + 2.19T_e^3) cm^3 s^{-1} \qquad (R2)$$ [15]

The rotational temperature $T_{rot}$ can be approximated [16] as the gas temperature $T_g$, i.e. $T_{rot} \approx T_g$ for partial local thermodynamic equilibrium (PLTE) plasmas. The production rate coefficient of $OH$ radicals of reaction $(R3)$ is, taking the typical value of the rotational temperature $T_g = 2200K$ as determined from the fitting of the $OH$ bands, $k_3 = 1.69 \times 10^{-10} cm^3 s^{-1}$. It is seen from reaction $(R3)$ and Fig. 3(a) that with increasing with $O_2$ flow rate the production of $O(\,^1D)$ is increased [17] hence more $OH(X)$ radicals are producing within the discharge gap. $e + OH(X) \rightarrow OH(A) + e,\ k = 2.7 \times 10^{-10} T_e^{0.5} \approx 2.06 \times 10^{-10} cm^{-3}s^{-1}$ is also an efficient channel for the production $OH(A)$ radicals[18].

$$O(\,^1D) + H_2O \rightarrow OH(X) + OH(X) \qquad k_3 = 1.62 \times 10^{-10} exp\left(\frac{64.95}{T_g}\right) cm^3s^{-1} \qquad (R3)^{[18]}$$

The destruction rate coefficients of $OH$ radicals through reactions $(R4)$ and $(R5)$ are $k_4 = 1.58 \times 10^{-11} cm^3s^{-1}$ and $k_5 = 2.07 \times 10^{-11} cm^3s^{-1}$, respectively, for the typical values of $T_g \approx 2200K$. Observations of reactions $(R4)$, $(R5)$, Figs. 3(a) and 3(b) reveal that the density of $OH$ radicals is decreasing with increasing $O_2$ flow rate because of lowering gas mixture temperature.

$$OH(X) + OH(X) \rightarrow O + H_2O \qquad k_4 = 2.5 \times 10^{-15} (T_g)^{1.14} exp\left(-\frac{50}{T_g}\right) cm^3s^{-1} \qquad (R4)^{[18]}$$

$$O(\,^1D) + OH(X) \rightarrow H + O_2 \qquad k_5 = 2.0 \times 10^{-11} T_g^{-0.186} exp\left(-\frac{154}{T_e}\right) cm^3s^{-1} \qquad (R5)^{[18]}$$

The loss mechanism of $OH(A)$ at high water concentration is mainly determined by quenching of $H_2O$ and $O_2$ [19] by the following reaction channels that appear in the discharge zone between the gap of the electrodes inside the discharge tube. The quenching of excited $OH(A)$ radicals with $H_2O$ and $O_2$ is approximately 5% [20]

$$OH(A) + H_2O \rightarrow OH(X) + H_2O \qquad k_6 = 4.9 \times 10^{-10} \left(\frac{T_g}{300}\right)^{0.5} cm^3s^{-1} \qquad (R6)^{[18]}$$

$$OH(A) + O_2 \rightarrow OH(X) + O_2(A) \qquad k_7 = 7.5 \times 10^{-11} \left(\frac{T_g}{300}\right)^{0.5} m^3s^{-1} \qquad (R7)^{[18]}$$

The quenching rate coefficients of reactions $(R6)$ and $(R7)$, estimated considering the same gas temperature $T_g = 2200K$, are $k_6 = 1.33 \times 10^{-9} cm^3s^{-1}$ and $k_7 = 2.03 \times 10^{-10} cm^3s^{-1}$, respectively. Due to addition of $O_2$, the rotational temperature is being lower from $2450K$ to $2000K$ at $5kV$ for the change of $O_2$ flow rate from 2 $lpm$ to 10 $lpm$ (as can be seen from Fig. 5(a)) and as a result quenching rate is also decreasing. It is to be noted that the quenching of rate of reaction $(R6)$ is higher than that of $(R7)$. In order to determine the ground state density of $OH$ radicals, laser induced fluorescence spectroscopic (LIF) technique is usually employed instead of using OES technique due to lack [9,11, 13] of direct relation between the intensity of the $OH(A)$-related emission bands and the ground state density of $OH$ radicals.

## 4. Result and Discussion

### 4.1 Rotational and vibrational temperatures

It is well known [21] that the emitted lines from the plasma are broadened by several broadening mechanisms. These are natural-, Doppler-, instrumental-, pressure- , resonance van der Waals- and Stark broadenings [22, 23]. Attention has been given to determine gas (rotational) temperature due to its key role for biological applications of $OH$ radicals. The species used for spectroscopic diagnostics that have significant contributions in the emitted spectra. The emission intensity of the species produced is dependent on the density of excited species, rotational and vibrational energy distributions, and the volume of the plasma under consideration. In our experiment, $OH(A-X)$ radicals produced highest intensity, indicating highest concentration, among other species as shown in Fig. 2. The emission spectra of $OH(A-X)$ radicals are widely used [9, 24] for spectroscopic diagnostics because it facilitates one to determine rotational ($T_{rot}$) and vibrational ($T_{vib}$) temperatures by fitting the measured spectra. LIFBASE software [25] is used for the determination of $T_{rot}$ and $T_{vib}$ because of its versatile applications. The emitted spectral line of Eq.(1) can be modified to the following form [9]

$$I_{n''v''J''}^{n'v'J'} = C(\nu_{n''v''J''}^{n'v'J'})^4 N_{n'} S_{v'v''} S_{J'J''} \times \frac{exp\left\{-\left(\frac{hc}{k}\right)\left(\frac{G_{n'}\left(v'\right)}{T_{vib}}\right)\right\}}{Q_{vib,n'}(T_{vib})} \frac{exp\left\{-\left(\frac{hc}{k}\right)\left(\frac{F_{v'}^{(i)}(J')}{T_{rot}}\right)\right\}}{Q_{rot,n'v'}^{(i)}(T_{rot})}, \quad (2)$$

where $C$ is a constant, $k$ is the Boltzmann constant, $c$ is the velocity of light, $N_{n'}$ is the density per volume of the molecules in the electronic state of $n'$, $S_{v'v''}$ is the band strength, $S_{J'J''}$ is the line strength, $G_{n'}\left(v'\right)$ is the harmonic oscillator term value, $F_{v'}^{(i)}(J')$ is the vibrational term value, $Q_{vib,n'}(T_{vib})$ is the vibrational partition function, and $Q_{rot,n'v'}^{(i)}(T_{rot})$ is the rotational partition function. The contribution of the Lorentzian function is taken into account considering the non-Gaussian [9] measured line shape. The best fitted line shape can be obtained due to the contributions of Lorentzian and Voigt profiles. Under these conditions, the broadened line intensity profiles can be represented [9] as a function of wavelength ($\lambda$)

$$I(\lambda) = I_{n''v''J''}^{n'v'J'}\left[1 - M\, exp\left\{\left(-\frac{\lambda - \lambda_{n''v''J''}^{n'v'J'}}{FWHM}\right)^2 4\ln(2)\right\} + M/4\left\{\left(\frac{\lambda - \lambda_{n''v''J''}^{n'v'J'}}{FWHM}\right)^2 + 1\right\}\right], \quad (3)$$

where $M$ is the fraction of the Lorentzian contribution to the Voigt profile of the instrumental function, $FWHM$ is the full width at half maximum and $\lambda_{n''v''J''}^{n'v'J'}$ is the $n'v'J' \rightarrow n''v''J''$ transition wavelength.

The rotational emission of $OH(A-X)$ radicals in the $306-318nm$ band is used to determine $T_{rot}$. Both the experimental and simulated spectra are normalized by $OH(0-0)$ band. The simulated spectrum is superimposed on the measured spectrum. The best match provides the $T_{rot}$ and $T_{vib}$ temperatures. The experimental data measured at a voltage of $5kV$, electrode spacing of $3mm$, and $O_2$ flow rate of $4\,lpm$ ($11.11\%\, H_2O$) along with best matched spectrum are shown in Fig. 4 for example. In this case, the instrumental broadening of $1.7nm$, Doppler broadening of $0.0028nm$, collisional broadening of $0.65nm$ are obtained from the simulated spectrum using LIFBASE software. It is noted that the LIFBASE software includes transition moment, rovibrational wavefunction, rotational Hönl-London factors and the emission coefficient with vibration and rotational quantum numbers that are essential parameters for the spectroscopic computations for the determination of accurate rotational and vibrational temperatures even with low resolution spectroscopic data [25].

Figure 5(a) and 5(b) show the effect of voltage and electrode spacing on $T_{rot}$ and $T_{vib}$ for different $O_2$ flow rate, respectively. It is observed from Fig. 5(a) that both $T_{rot}$ and $T_{vib}$ are increasing with the increase of voltage with enhanced $O_2$ flow rate for the electrode spacing of $3\,mm$. It is noted that the flow of $H_20$ steam was kept constant for all measurements. The reasons of the increasing $T_{rot}$ and $T_{vib}$ arise due to the fact that with increasing voltage, free electrons are collecting more energy from the increased electric field and transferring this energy to neutral particles through electron-neutral moment transfer process. Fig. 5(b) displays the effect of electrode spacing on $T_{rot}$ and $T_{vib}$ maintaining the discharge voltage of $5\,kV$. It is seen from this figure that $T_{rot}$ and $T_{vib}$ are decreasing with increasing electrode spacing. This result can be attributed from the fact that for a given applied voltage and atmospheric pressure, as the electrode spacing is increased, the electrical field intensity across the electrodes are decreasing, therefore, the mean energy of the electron is reduced and hence less energy from the electrons are transferred to the atoms or molecules. It is also observed from Fig. 5 that $T_{rot}$ and $T_{vib}$ are reducing with the increase of $O_2$ flow rate due to the fact that, $O_2$ plays a significant role in carrying Joule heat [26]out from the discharge region. It is to be mentioned that the fraction of energy is transferred by electron to $H_2O$ and $O_2$ molecules by vibrations and rotations. But according to our experimental conditions, there is no significant effect of $O_2$ flow rate and mixture temperature on high E/N values [26]. When $O_2$ flow rate is increased, the collision between molecules is more frequent due to short collision path length. This can be attributed by the fact that the Reynolds number is increasing with increasing gas flow rate in the electrode gap region. The diameter of the discharge tube are different at the entrance ($10mm$) and exit ($2mm$). So it is reasonable to consider that the gas molecules will be accumulated in the discharge region with increasing $O_2$ flow rate. Because we can represent the Reynolds number as a function of gas density, the Reynolds number $R_e = n_g v_g l_c / \mu_d$, where $n_g$, $v_g$, $l_c$, and $\mu_d$ are the gas density, gas velocity, characteristic length and dynamic viscosity, respectively. The results obtained agree well with results noted by Zhang *et al.* [27]. Due to high collision frequency, the electrons are unable to gain much energy from the applied electric field and hence transfer less energy to molecules through momentum transfer collision process. On the other hand, the electron transit time becomes shorter within the electrode gap where strong electric field exists. Anghel *et al.* [28] investigated the effect of operating frequency (MHz range) and gas flow rate ($1 - 5\ lpm$) on plasma power dissipation and revealed that the change of power dissipation is very small with operating frequency and gas flow rate. But in our experiment the source frequency is 88 $Hz$ only, and the effect of discharge power variation can be considered negligible with gas flow rate and frequency.

**4.2 Electron Density**

Recently, the interest for the spectroscopic diagnostics of atmospheric pressure using $H_\alpha$ line is considerably increased. The intense intensity of $H_\alpha$ line appears when only traces of hydrogen and $H_2 0$ are present in the plasma [29]. In this experiment, the spectroscopic measurements show that the $H_\alpha$ lines produced highest intensity instead of $H_\beta$ line and hence the $H_\alpha$ lines are considered for the plasma diagnosis. Doppler broadening ($\lambda_D$) is produced due to the thermal motion of excited hydrogen atoms. The instrumental broadening ($\lambda_i$) is related to the resolution of the spectrometer, slit function and lens error of the optical transmission system. The pressure broadening includes three broadening mechanisms: the Stark broadening ($\lambda_s$), (due to the interaction of charged particles in plasmas)[30], the resonance broadening ($\lambda_r$) and Van der Waals broadening ($\lambda_{van}$). According to our experimental conditions, the natural broadening ($\lambda_n$) and resonance broadening ($\lambda_r$) are usually negligible [30] in most plasma experiment because resonance broadening does not induce shift of spectral line in low density plasmas.

However, in order to determine electron density, the spectral line plays an important role to obtain broadening parameters. The broadening mechanisms can be represented by the Gaussian and Lorentzian line profiles. The FWHM of the Gaussian profile $\Delta\lambda_G$ is given by [31, 32]

$$\Delta\lambda_G = \sqrt{{\lambda_D}^2 + {\lambda_i}^2}\ , \qquad (4)$$

while the total FWHM $\Delta\lambda_L$ is directly the sum of the each individual contribution.

$$\Delta\lambda_L = \lambda_n + \lambda_r + \lambda_{van} + \lambda_s\ , \qquad (5)$$

The experimental spectrum of the $H_\alpha$ line is fitted by Voigt profile which is the convolution of $\Delta\lambda_G$ and $\Delta\lambda_L$. Since the line shapes contain all broadening mechanisms involved [33], hence all broadening parameters were included in the fitting equation as noted in the LIFBASE software. The Voigt fit of $H_\alpha$ line shape at wavelength $656.34nm$ is shown in Fig. 6, measured at the applied voltage of $5kV$ with electrode spacing of $3mm$ and $O_2$ flow rate of $4lpm$ ($11.11\%\ H_2O$). To estimate the value of the broadening, the best fitting are considered, excluding error by setting three important parameters: the central value, relative intensity and the base line of the spectrum. It is noted that $\lambda_n$ and $\lambda_r$ can be neglected [30, 31] from the contribution of the Lorentzian profile due to low resolution of the spectrometer used. But the van der Waals $\lambda_{van}$ broadening is usually accounts from 30% to 35% of the total FWHM of Lorentzian profile [31]. Under the experimental conditions considered, the typical van der Waals broadening $\lambda_{van} = 0.69nm$ is obtained. Using Eq. (5) the Stark broadening $\lambda_S$ can be obtained from which electron density is determined from the following equation [21, 22]

$$\lambda_S = 2.0 \times 10^{-11} {n_e}^{2/3}\ cm^{-3} \qquad (6)$$

Figures 7(a) and 7(b) show the dependency of electron density $n_e$ on the applied voltage and electrode spacing for different $O_2$ flow rate, respectively. Figure 7(a) indicates that $n_e$ decreases with increasing electrode spacing while decreases with increasing $O_2$ flow rate. Electric field intensity within the gap reduces with increasing electrode spacing and hence $n_e$ decreases. It is well known that at intense electric field, the electrons gain more energy from the enhanced electric fields and hence the ionizing collisions between electrons and neutrals become more frequent. Alternatively, the higher the value of electron temperature, the more effectively ionizes the neutrals causing abundant production of electrons. Therefore, most of the energies of the electrons, as gained from the electric field, being transferred to the $O_2$ species within the electrode gap through ohmic heating [33] and thereby producing more electrons and consequently $n_e$ increases with increasing voltage at constant $O_2$ flow rate. Figure 7(b) shows that $n_e$ is decreased with increasing $O_2$ flow rate. As increasing $O_2$ flow rate, $n_e$ is decreasing due to the decrease of water molecule content in discharge region, because of constant $H_2O$ flow, and high electron affinity of the $O_2$ [33] molecules. With increasing $O_2$ flow rate, the collision of electrons with $O_2$ become frequent and hence $n_e$ decreases with a small amount of $H_2O$ molecule content because of electron quenching and the electronegativity nature of $O_2$ molecules.

### 4.3 Excitation temperature $T_{ex}$

The excitation temperature ($T_{ex}$) may be used as a rough indicator of the electron temperature [34]. Because the free electrons are responsible for the excitation of atoms or molecules and their energies should be described by the Boltzmann distribution function at a given temperature. Intensity ratio method is the most straightforward method to determine the excitation temperature from the line emission if the population densities of the upper levels of two lines are in partial local thermodynamic equilibrium [35]. The intensity of the emission lines provides the population in the level of atoms that follow the Boltzmann distribution function the relative intensity of two different lines at wavelength $\lambda_1$ and $\lambda_2$ can be expressed by following relation [36, 37]

$$\frac{I_1}{I_2} = \frac{A_1 g_1 \lambda_2}{A_2 g_2 \lambda_1} exp\left(-\frac{E_2 - E_1}{kT_e}\right), \qquad (7)$$

where $I$ refers to the intensities of the emission lines, $A$ refers the transition probabilities, $g$ is the degeneracy of the upper levels, $E$ is the upper level energies and $k$ is the Boltzmann constant. Exponential function in Eq. (7) reveals that the lines should be selected in such a way that the condition $kT_e < E_2 - E_1$ satisfies. In the present analysis, $T_{ex}$ is determined from the atomic transitions of $CuI$ lines at the wavelength of $324.12\,nm$ and $327.32\,nm$, respectively as shown in Fig. 2. The intensities of the $CuI$ lines are taken from the measured spectrum and the relevant spectroscopic data for the transitions considered are taken from NIST database [38]. Figures 8(a) and 8(b) show the effects of $O_2$ flow rate and electrode spacing on $T_{ex}$. It is seen from Fig. 8(a) that the $T_{ex}$ is increasing with the increase of $O_2$ flow rate and applied voltage. The reason of increasing $T_{exc}$ arises due to the fact that the presence of electronegative $O_2$ gas in the plasma causes the increase of sustaining voltage and the electric field strength. Therefore, the mean kinetic energy is increased and lead to an increase of the excitation temperature. In this range, the contribution of penning ionization decreases due to the reduction of the available $OH$ species in the plasmas [39]. The observation of $T_{exc}$ is in agreement with the results of Joh *et al.*[34]. With increasing electrode spacing the mean kinetic energy of electrons are reduced, as shown in Fig. 8(b), as discussed earlier.

**5. Conclusion**

The atmospheric pressure $H_2O/O_2$ gliding arc plasma is produced. The optical emission spectroscopic data as recorded from the plasma discharges reveals that the emission band of $OH(X^2\Pi(v' = 0) \rightarrow A^2\sum^+ v'' = 0))$ and atomic transition of $O, H_\alpha and\ H_\beta$ are produced. The plasma parameters $T_{rot}$, $T_{vib}$, $T_{exc}$ and $n_e$ are determined by analysing optical emission spectroscopic data in the voltage range from $4.0 - 5.5kV$ and the electrode spacing from $2.0 - 5.0\,mm$. The analytical results show that $T_{exc} > T_{vib} > T_{rot}$ which means that the plasma discharges are in non-thermodynamic equilibrium. $T_{rot}$, $T_{vib}$, $T_{exc}$ and $n_e$ are increased with increasing applied voltage but decreasing with electrode spacing. On the other hand, $T_{rot}$ and $T_{vib}$ are decreased with increasing $O_2$ flow rate while $T_{exc}$ and $n_e$ are decreased.

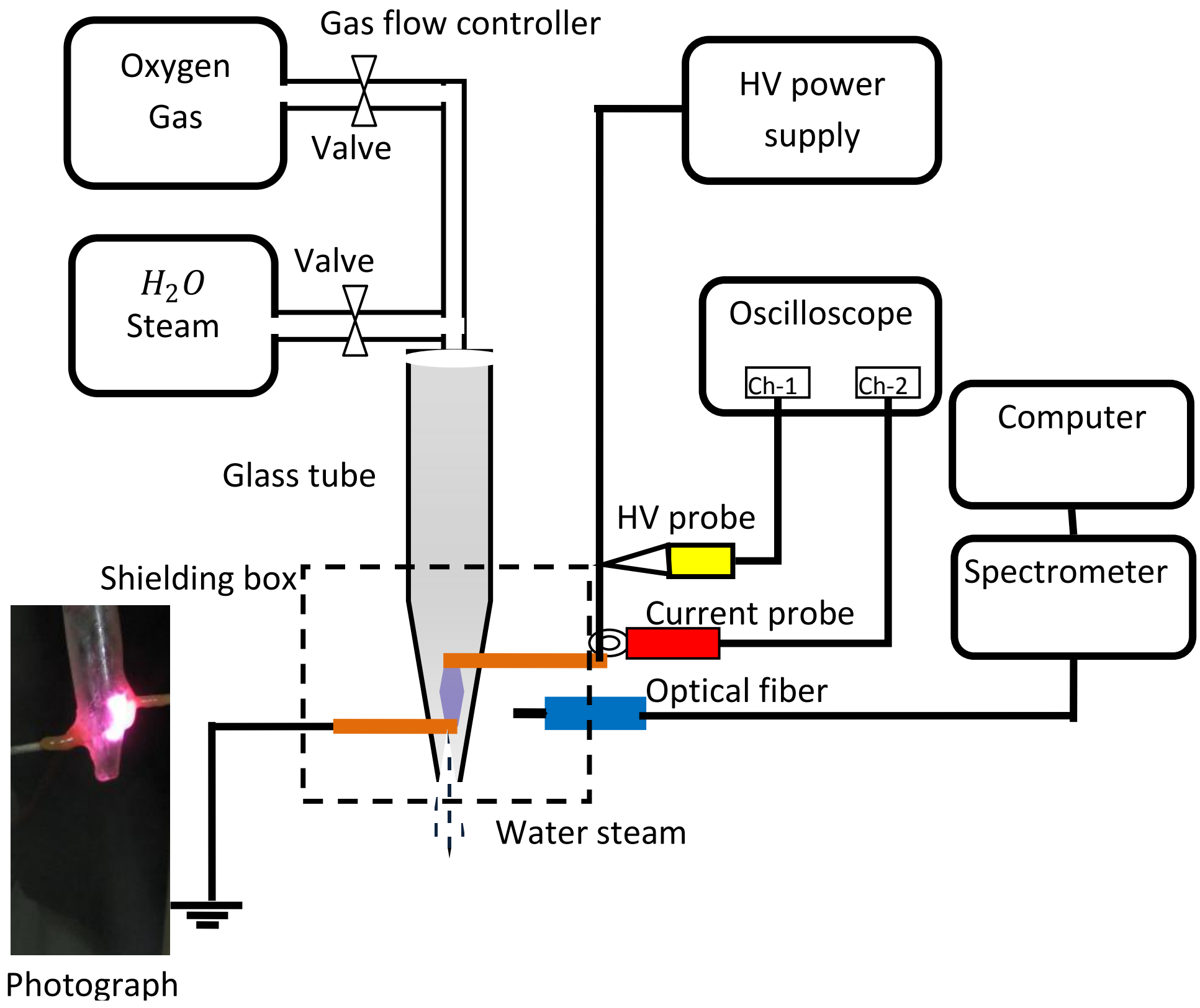

**Fig. 1.** Schematic of the experimental setup with discharge photograph

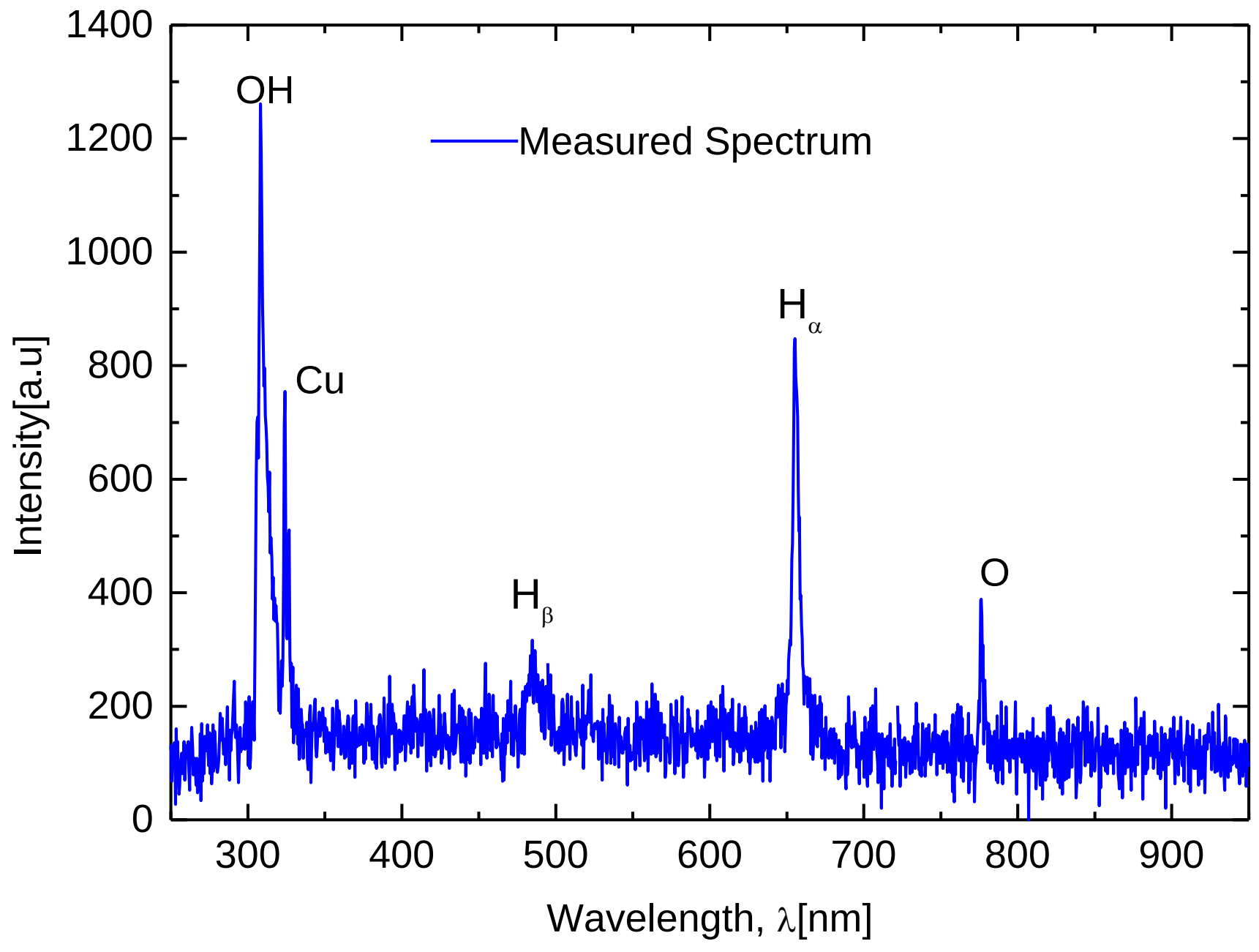

**Fig. 2.** Emission spectra of steam with $O_2$ flow rate of $4lpm$ ($11.11\%\ H_2O$), voltage $5kV$, and electrode gap $3mm$.

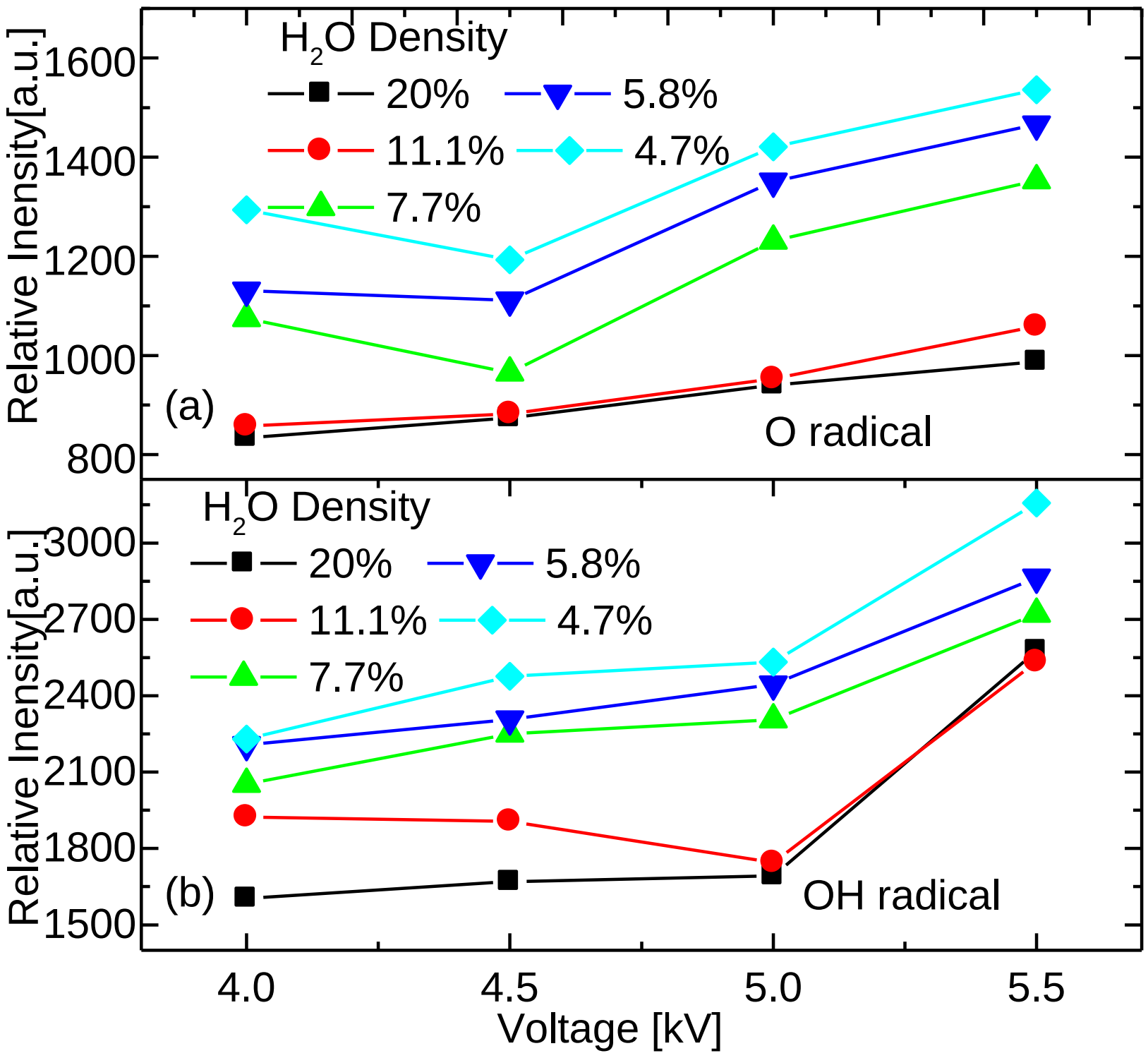

**Fig. 3.** Relative intensity of $O$ and $OH(A-X)$ radicals with voltage for different $H_2O$ content for $3mm$ electrode spacing.

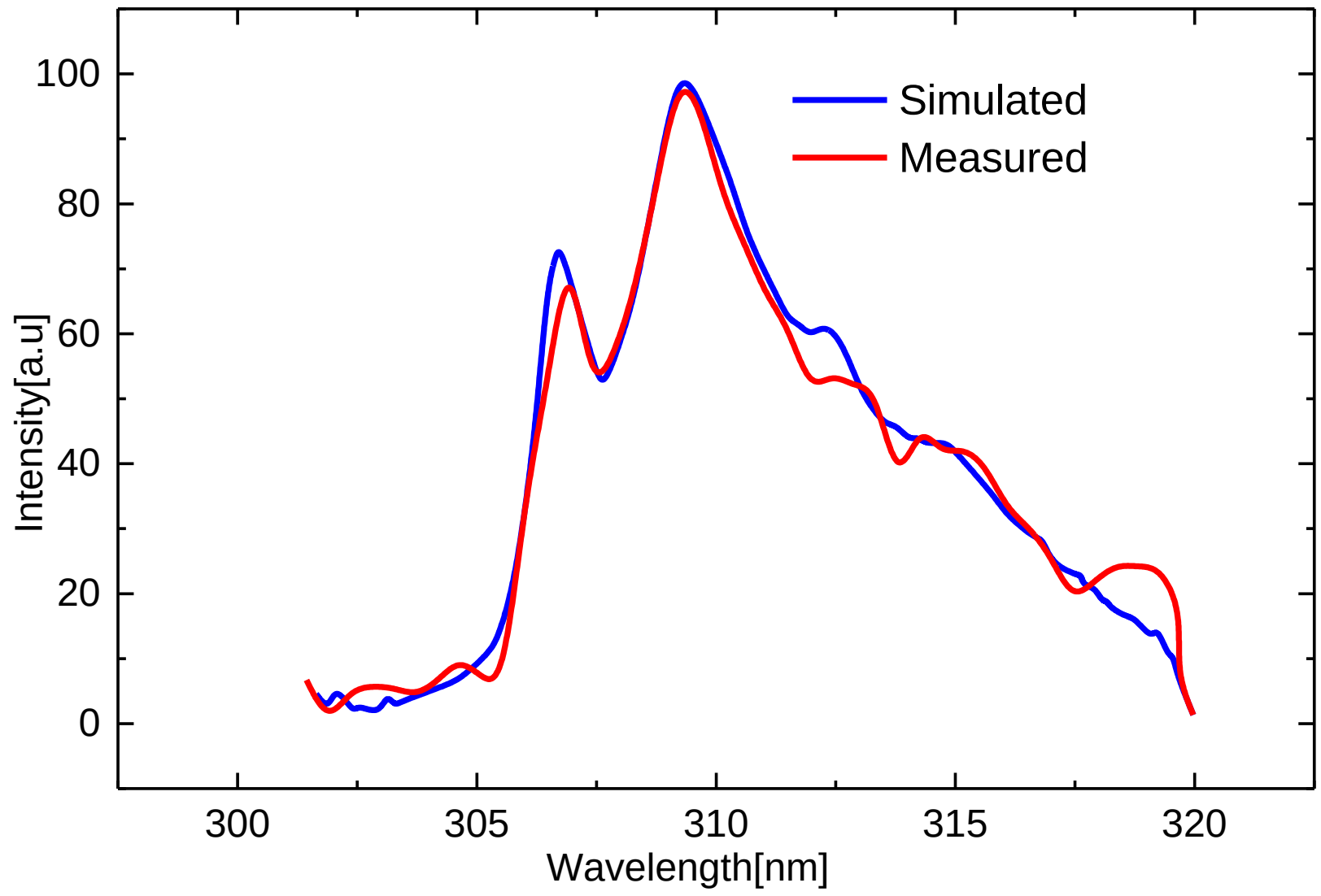

**Fig. 4.** Experimental and simulated spectra of $OH(X^2\Pi(v^{'}=0) \rightarrow A^2\sum^{+} v^{''}=0))$ radical emitted from $H_2O$ with $O_2$ flow rate of $4lpm$ ($11.11\%\ H_2O$), voltage of $5kV$ and electrode spacing of $3mm$.

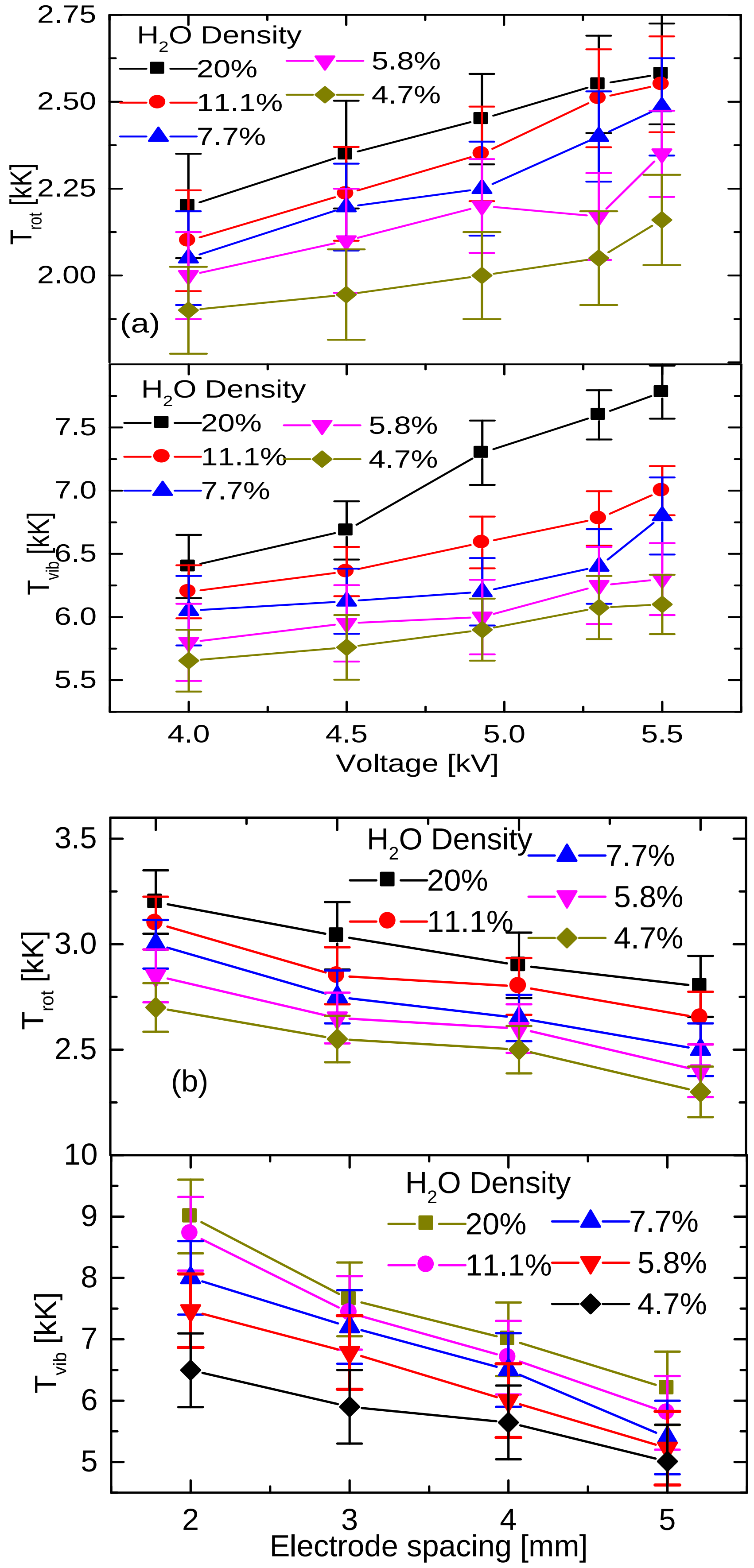

**Fig. 5.** Effect on rotational ($T_{rot}$) and vibrational ($T_{vib}$) temperatures (a) of voltage determined at $3mm$ electrode spacing, and (b) of electrode spacing determined at 5 $kV$.

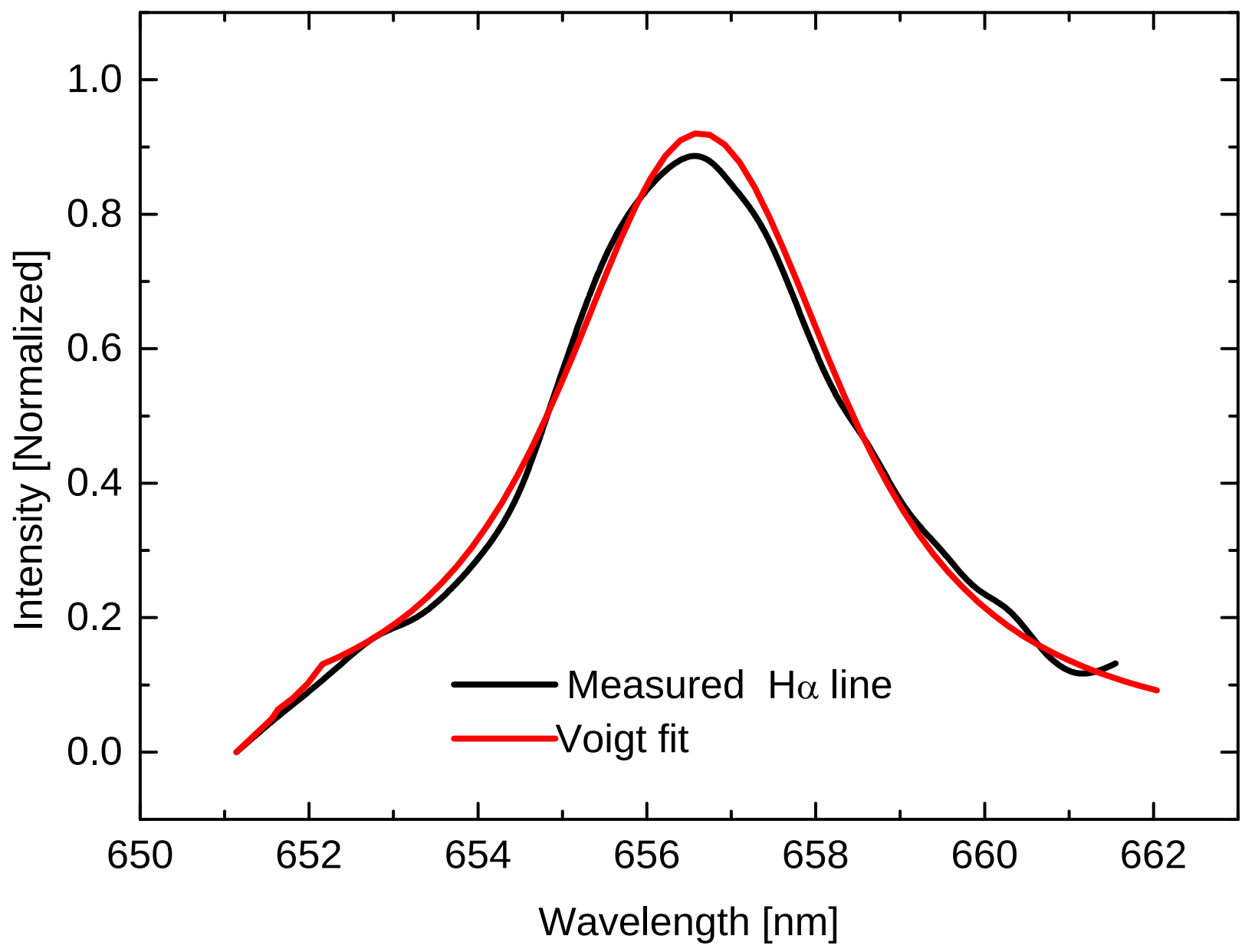

**Fig. 6**. Measured ($O_2$ flow rate 4 $lpm$ (11.11% $H_2O$), electrode spacing $3mm$ and voltage $5kV$) and fitted curves of $H_\alpha$ line for the determination of electron density from the broadening parameters.

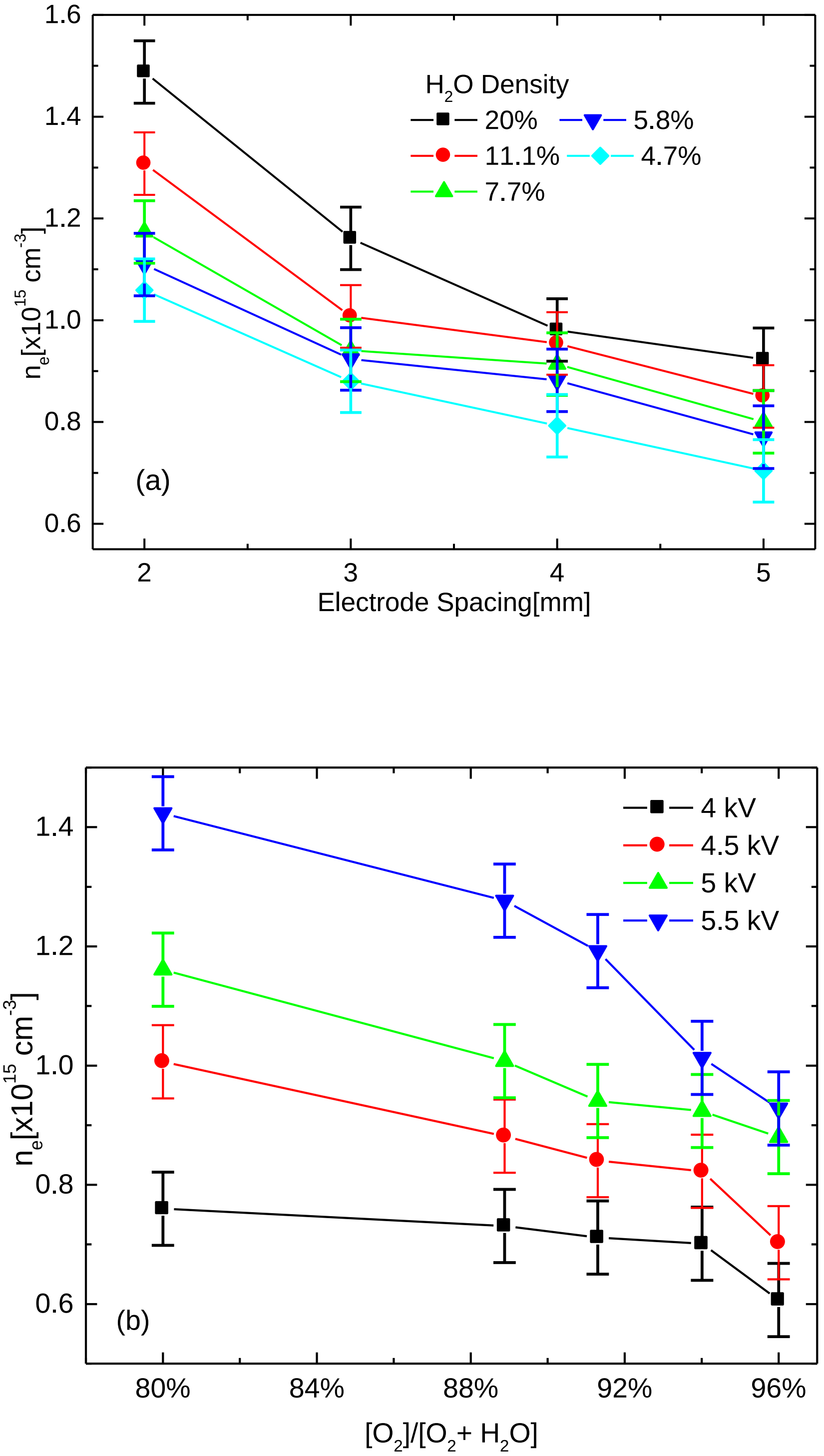

**Fig. 7.** Effect on electron density $(n_e)$ of (a) electrode spacing at $5kV$ and (b) $[O_2]/[O_2 + H_2O]$ with electrode spacing of 3 $mm$.

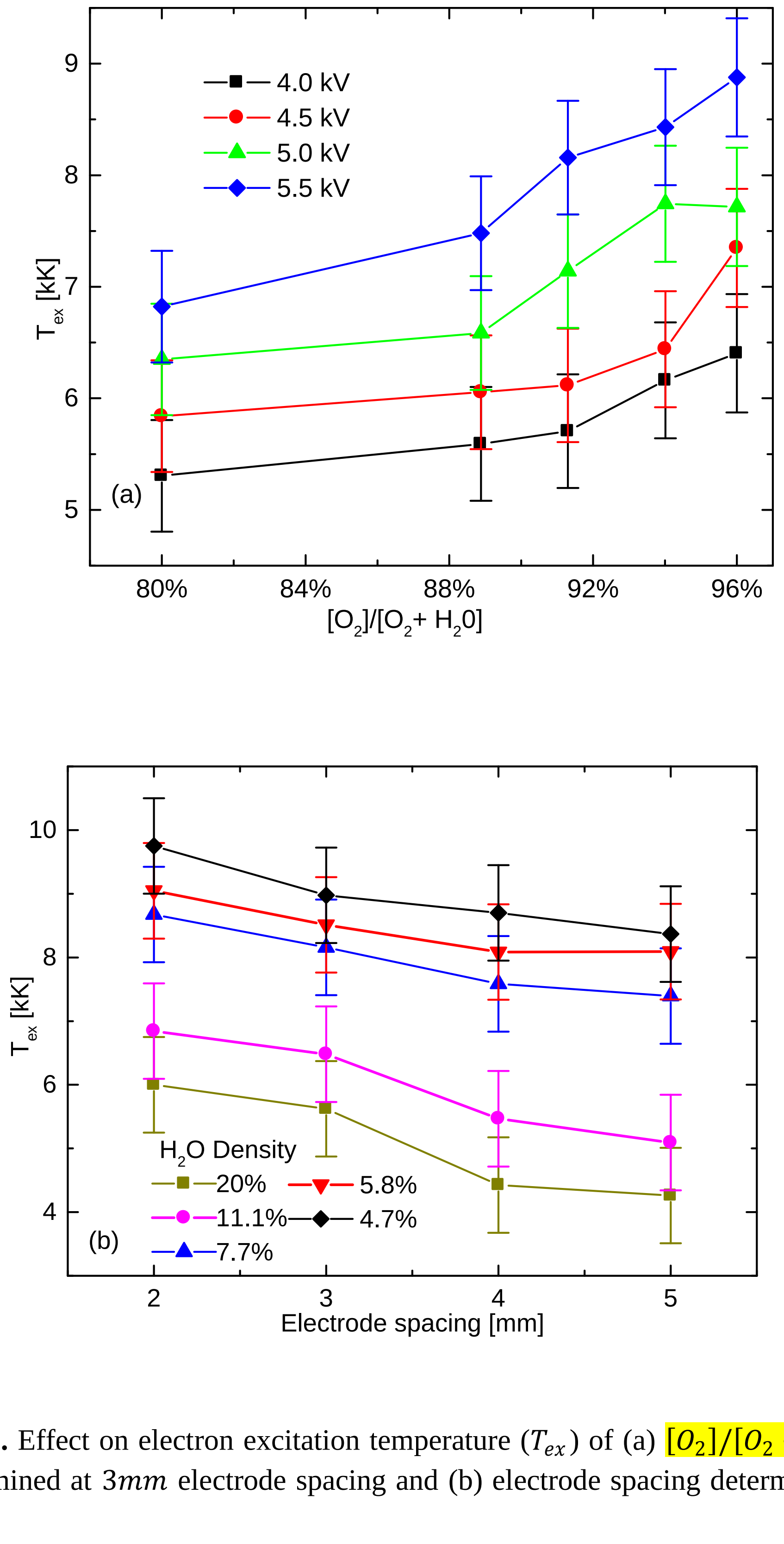

**Fig. 8.** Effect on electron excitation temperature ($T_{ex}$) of (a) $[O_2]/[O_2 + H_2O]$ determined at $3mm$ electrode spacing and (b) electrode spacing determined at $5kV$.